\documentclass[%
 reprint,
superscriptaddress,
 amsmath,amssymb,
 aps,
]{revtex4-2}

\usepackage{graphicx}
\usepackage{dcolumn}
\usepackage{bm}
\usepackage{hyperref}
\usepackage{comment}
\usepackage{lineno}
\usepackage{enumitem}
\usepackage{wrapfig}
\usepackage{tikz}

\tikzset{snake it/.style={decorate, decoration=snake}}

\begin{document}

\preprint{APS/123-QED}

\title{
Revisiting Thermal Charge Carrier Refractive Noise in Semiconductor Optics for Gravitational-Wave Interferometers}

\author{Harrison Siegel}
\email{hs3152@columbia.edu}
\affiliation{
 Department of Physics, Columbia University,
704 Pupin Hall, 538 West 120th Street, New York, New York 10027, USA
}
\affiliation{
Center for Computational Astrophysics, Flatiron Institute,
162 5th Avenue, New York, New York 10010, USA}
\author{Yuri Levin}%
 \email{ylevin@flatironinstitute.org}
\affiliation{
 Department of Physics, Columbia University,
704 Pupin Hall, 538 West 120th Street, New York, New York 10027, USA
}
\affiliation{
Center for Computational Astrophysics, Flatiron Institute,
162 5th Avenue, New York, New York 10010, USA}
\affiliation{Department of Physics and Astronomy, Monash University, Clayton, VIC 3800, Australia}


\newcommand{\RomanNumeralCaps}[1]
    {\MakeUppercase{\romannumeral #1}}
    
\begin{abstract}
The test masses in next-generation gravitational-wave interferometers may have a semiconductor substrate, most likely silicon. The stochastic motion of charge carriers within the semiconductor will cause random fluctuations in the material's index of refraction, introducing a noise source called Thermal Charge Carrier Refractive (TCCR) noise. TCCR noise was previously studied in 2020 by Bruns et al., using a Langevin force approach. Here we compute the power spectral density of TCCR noise by both using the Fluctuation-Dissipation theorem (FDT) and accounting for previously neglected effects of the standing wave of laser light which is produced inside the input test mass by its high-reflecting coatings. We quantify our results with parameters from Einstein Telescope, and show that at temperatures of 10~K the amplitude of TCCR noise is up to a factor of $\sqrt{2}$ times greater than what was previously claimed, and from 77~K to 300~K the amplitude is around 5 to 7 orders of magnitude lower than previously claimed when we choose to neglect the standing wave, and is up to a factor of 6 times lower if the standing wave is included. Despite these differences, we still conclude like Bruns et al. that TCCR noise should not be a limiting noise source for next-generation gravitational-wave interferometers.

\end{abstract}

\maketitle


\section{introduction}
The current second generation of gravitational-wave intereferometer experiments such as LIGO \cite{AdvancedLIGOScientific:2014pky}, Virgo \cite{VIRGO:2014yos}, and Kagra \cite{Kagra_ref}, has been highly successful in measuring the gravitational effects of astrophysical compact object mergers, and has thus ushered in a new and exciting age of multi-messenger astronomy. Proposed designs for next-generation gravitational-wave interferometer experiments including LIGO Voyager \cite{LIGOvoyager}, Einstein Telescope (ET) \cite{ETDesignReport}, and Cosmic Explorer (CE) \cite{CE_Horizon_study}, aim to measure gravitational strains with up to a full order of magnitude better sensitivity than that of the current generation, and thus open the door to observations of many more compact object mergers in addition to other astrophysical phenomena whose signatures are too weak to be detected by current experiments. The next generation's increase in precision will be achieved by a variety of improvements such as longer interferometer arms, new laser technologies, implementation of both L-shaped (Voyager, CE) and triangular (ET) interferometer geometries, and new test masses which may replace the currently used fused silica substrate with a semiconductor material and may also be cooled to cryogenic temperatures.

The motivation for switching from fused silica to semiconductor test masses is the mitigation of thermal noise, a prominent limiting noise source. The most likely choice of semiconductor material is silicon, since the thermal properties of silicon compare favorably to those of fused silica in many respects: silicon has lower Brownian noise than fused silica when cooled to cryogenic temperatures; fused silica mirrors undergo significant thermal deformation when impurities in the test mass coatings absorb laser power, whereas silicon suppresses these deformations due to its high thermal conductivity; and at temperatures of 18 K and 125 K, silicon's thermoelastic noise vanishes because the thermal expansion coefficient goes to zero \cite{CE_Horizon_study, ETDesignReport, Si_thermalexpansion_paper}.

However, semiconductor test masses will also give rise to new noise sources. In particular, semiconductors, unlike fused silica, possess a conduction band which allows free charge carriers to move throughout the material, and this introduces a new noise source known as Thermal Charge Carrier Refractive (TCCR) noise which will be the focus of our paper. TCCR noise is produced by time-dependent thermal fluctuations of charge carrier density within the semiconductor material which in turn cause local variations of the index of refraction, thus changing the overall phase of laser light passing through the input test mass. TCCR noise was recently investigated by Bruns et al. \cite{TCCRPaper}, and is similar to the thermochemical noise described by Benthem and Levin for GEO600 \cite{BenthemLevin}.

The previous TCCR noise computation performed by Bruns et al. makes use of Langevin forces to model Brownian motion of the charge carriers. It assumes that the Langevin force $F_L(\vec{r},t)$ is an uncorrelated white Gaussian noise process described in Fourier space by
\begin{equation}
\begin{split}
    \langle F_L(\vec{k},w)F_L^*(\vec{k^\prime},w^\prime)\rangle = (2\pi)^4 F_0^2(k^2+l_D^{-2}) \\ \times\delta(\vec{k}-\vec{k^\prime})\delta(\omega-\omega^\prime),
\end{split}
\end{equation}  
where $F_0$ is a constant constrained by a requirement that charge carrier density fluctuations are a Poisson process, and $l_D$ is the Debye Length. The previous work also uses a laser beam form factor (an expression which is determined by the beam's intensity profile, and written in Cartesian coordinates where $z$ is the direction of beam propagation),
\begin{equation}
    q(\vec{r})=\frac{1}{\pi r_{0}^{2}} ~ \text{exp}\left[\frac{-(x^{2}+y^{2})}{r_{0}^{2}}\right],
    \label{intro_form_factor}
\end{equation}
where $r_0$ is the beam waist radius. This form factor notably only considers transverse components of the beam, ignoring the standing wave of laser light which is produced along the $z$ direction by reflections off the high-reflecting coatings located on the furthest side of the input test mass from the beamsplitter.

We argue that a better approach than the Langevin method is a more general computation which uses the Fluctuation-Dissipation theorem (FDT) as was done in Benthem and Levin's thermochemical noise analysis, and we also claim that the standing wave in the form factor is significant and must be included. We find that our approach produces a different result than that of Bruns et al., and this is to be expected for the following reasons. The assumption that the Langevin forces have a Poisson process constraint may not be valid in the context of charge carrier density fluctuations, since a Poisson process consists of random probability throws which are independent whereas charge carrier motions are dependent on each other due to electrical repulsion and/or attraction. As for the neglected standing wave, its field gradient is much steeper than that of the transverse beam component since the length scales of both are set respectively by the laser light's wavelength and the comparatively large beam waist radius $r_0$. Due to the steeper field gradient, diffusion of charged particles over the standing wave should contribute much more to TCCR noise at higher temperatures where the Debye length starts to shrink to scales below the light's wavelength. It is also reasonable to expect suppression of the transverse noise contribution at these same temperatures, since the Debye length becomes so much smaller than $r_0$ that the spatial extent of any given density fluctuation essentially does not overlap with different regions of transverse beam intensity.

In this paper we derive the power spectral density of TCCR noise by using the FDT and considering a wide range of temperatures and semiconductor doping concentrations. We quantify our results using parameters from the ET low-frequency detector. Our results differ from those of Bruns et al. both at low and high temperatures. At temperatures around 10 K, the FDT-derived noise amplitude is up to a factor of roughly $\sqrt{2}$ times greater than what is obtained by using the Langevin method. Between 77~K and 300~K, the standing wave becomes extremely significant. When neglecting the standing wave at these temperatures, depending on the doping concentration the noise amplitude is 5 to 7 orders of magnitude lower than what was previously claimed, and when including the standing wave the amplitude is up to 6 times smaller than previous claims.

\section{FDT for TCCR Noise: Theory}
Here we broadly follow \cite{BenthemLevin}, which is based on the FDT as formulated in \cite{PhysRev.83.34} and \cite{Levin1998}. The FDT gives an expression for the power spectral density $S_Q$ of any generalized coordinate $Q$ of a linear dissipative system. In particular, if $Q$ is being acted on by a generalized driving force $F=F_0\text{exp}(i\omega t)$, then $S_Q$ can conveniently be written in terms of $W_{\rm diss}$, the power dissipation from $F$ time-averaged over one period, as follows:
\begin{equation}
    S_Q(f)=\frac{8 k_B T}{\omega^2}\frac{W_{\rm diss}}{F_0^2},
    \label{S_Q_general}
\end{equation}
where $T$ is the temperature of the system, $f=\omega/(2\pi)$ is the frequency, and $k_B$ is the Boltzmann constant. Note that the above expression is obtained by taking the classical limit, $k_BT >> \hbar\omega$. {\bf Importantly}, in the above $F$ is written as a phasor, and therefore we implicitly take the real part of all measurable quantities linearly related to $F$ throughout this paper.

We can use the FDT to compute the power spectral density of TCCR noise by defining our generalized coordinate to be the experimental readout variable $\delta z$, the optical path length change resulting from time-dependent fluctuations of free charge carrier density along the path of the laser light as it passes through the input test mass. This readout variable can be expressed as 
\begin{equation}
    \delta z (t) = \int_{V}d^3\vec{r}~ \alpha ~ \delta N(\vec{r}, t) ~   q(\vec{r}),
\end{equation}
where $V$ is the input test mass volume, $\alpha=\partial n /\partial N$ where $n$ is the index of refraction and $N$ is the number density of charge carriers, $\delta N(\vec{r} , t)$ is the relative change in number density of charge carriers along the light's path,  and $q(\vec{r})$ is the beam form factor. We note that as in \cite{TCCRPaper},
\begin{equation}
    \alpha = \frac{-e^2}{2 n \epsilon_0  m_e \omega_l^2 },
\end{equation}
where $e$ is the electric charge of a  carrier (throughout this paper, without loss of generality we consider a semiconductor with majority charge carriers which are electrons), $\epsilon_0 $ is the vacuum permittivity, $\omega_l$ is the laser light frequency, and $m_e$ is the effective mass of majority charge carriers.

As in \cite{BenthemLevin,ModernClassicalPhysics}, in order to find a generalized time-periodic force ${F}$ that drives the canonically conjugate momentum of $\delta z$, we introduce an interaction Hamiltonian  $H_{\text{int}}$:
\begin{equation}
    H_{\text{int}} = -F_{0} ~ \text{exp}(i\omega t) ~ \delta z.
    \label{H_int}
\end{equation}
Under the action of $H_{\text{int}}$, each charge carrier would experience a force $\vec{F}(\vec{r}_i,t)$ at its position $\vec{r}_i$:
\begin{equation}
\begin{split}
    \vec{F}(\vec{r}_i,t) & =  -  \nabla_i H_{\text{int}} \\
    & = F_{0} ~ \text{exp}(i\omega t) ~ \alpha ~ \nabla q(\vec{r})\Big|_{\vec{r}=\vec{r}_i}. 
    \label{generalized_force}
\end{split}
\end{equation}
In order to compute the noise power spectral density $S_{\delta z}$, we need to find the time-averaged power $W_{\rm diss}$ that would be dissipated under the action of the force field in Eq.~(\ref{generalized_force}). In the rest of this section we explain how to obtain $W_{\rm diss}$.

First we note that the force field in Eq.~(\ref{generalized_force}) would induce an oscillating current $\vec{j}(\vec{r},t)$ of charge carriers within the semiconductor substrate of the input test mass, along with charge density perturbations $\delta\rho(\vec{r},t)$ and associated electric fields $\vec{E}(\vec{r},t)$. For small charge density perturbations, the current $\vec{j}$ satisfies the following equation:
\begin{equation}
    \vec{j}=\rho_0\mu (\vec{F}+e\vec{E})-D\nabla\delta\rho
    \label{current_density},
\end{equation}
where $\mu$ is the generalized mobility, $\rho_0$ is the background charge density, and $D$ is the diffusion coefficient. Additionally, $\vec{j}$, $\vec{E}$, and $\delta\rho$ are related by the continuity equation and Gauss' law,
\begin{eqnarray}
\nabla\cdot \vec{j} &=& -\partial \delta \rho /\partial t\nonumber\\
\nabla\cdot \vec{E}& = & \delta\rho/(\epsilon\epsilon_0).\label{continuity}
\end{eqnarray}

By taking the divergence of Eq.~(\ref{current_density}) and using Eqs.~(\ref{generalized_force}) and (\ref{continuity}), we get
\begin{equation}
    \left(\frac{\partial}{\partial t}+\frac{D}{ l_D^2} -D\nabla^2\right)\delta\rho=-\rho_0\mu \alpha F_{0} ~ \text{exp}(i\omega t) ~ \nabla^2 q.
    \label{deltarho1}
\end{equation}
Here $l_D$ is the Debye length, the characteristic length over which free charges are screened inside the medium:
\begin{equation}
    l_D=\sqrt{\frac{\epsilon \epsilon_0 D}{ \rho_0\mu e}}.
\end{equation}
Equation (\ref{deltarho1}) has the general solution
\begin{equation}
    \delta\rho(\vec{r},t)=\delta\rho_0(\vec{r}) ~ \text{exp}(i\omega t).
\end{equation}
We can obtain an expression for $\delta\rho_0(\vec{r})$ in the Fourier domain, which we will use later:
\begin{equation}
    \tilde{\delta \rho}_0(\vec{k})=\frac{\rho_0\mu \alpha F_0 k^2}{ i\omega +{D / l_D^2}+D k^2}~\tilde{q}(\vec{k})
    \label{Fourier1}
\end{equation}
We shall be using the following convention for Fourier transforms:
\begin{eqnarray}
q(\vec{r})&=&\frac{1}{ (2\pi)^3}\int \tilde{q}(\vec{k})~e^{i\vec{k}\cdot\vec{r}}~d^3k
\nonumber\\
\tilde{q}(\vec{k})&=&\int q(\vec{r})~e^{-i\vec{k}\cdot\vec{r}} ~d^3r
\end{eqnarray}

We now have everything we need to compute $W_{\rm diss}$. First we will find the instantaneously dissipated power,
\begin{eqnarray}
W_\text{inst} &=&\int d^3r~\vec{F}\cdot\vec{j}/e\nonumber\\
  &=&\frac{F_0 \text{exp}(i\omega t)\alpha}{ e}\int d^3 r~\nabla q\cdot \vec{j}\label{intrelation}\\
  &=&\frac{F_0 \text{exp}(i\omega t)\alpha}{ e}\int d^3 r~ q ~\frac{\partial{\delta \rho}}{ \partial t}.\nonumber
\end{eqnarray}
The last step used $\nabla q\cdot \vec{j}=\nabla\cdot (q\vec{j})+q(\partial\delta\rho/\partial t)$ and the assumption that $q\vec{j}=0$ at the boundary of the integration domain. After applying the Plancherel theorem to Eq.~(\ref{intrelation}) to convert the integral into an expression in the Fourier domain and substituting in Eq. (\ref{Fourier1}), and then time-averaging (note that for any two phasors $A$ and $B$, taking 1/2~$\mathfrak{Re}(AB^*)$ gives the time average), we get
\begin{equation}
    W_{\rm diss}=\frac{F_0^2\alpha^2 N_0 \mu\omega^2}{ 16\pi^3}\int d^3k \frac{{k^2}\left|\tilde{q}(\vec{k})\right|^2}{ \omega^2+\left({D / l_D^2}+Dk^2\right)^2},
    \label{Wdiss2}
\end{equation}
where $N_0$ is the background number density of charge carriers. Substituting Eq.~(\ref{Wdiss2}) into Eq.~(\ref{S_Q_general}), we obtain
\begin{equation}
    S_{\delta z}(f)=\frac{D\alpha^2 N_0}{ 2\pi^3}\int d^3k \frac{{k^2}\left|\tilde{q}(\vec{k})\right|^2}{ \omega^2+\left({D / l_D^2}+Dk^2\right)^2},
    \label{PSDGeneral}
\end{equation}
where we used the Einstein relation $D=\mu k_B T$. This is the general expression for the power spectral density of TCCR noise. 

As in \cite{TCCRPaper}, in order to relate the change in optical path length to measured changes in gravitational strain within a given interferometer, we must multiply $S_{\delta z}$ by $\pi^2/(2 F_{\text{FP}}^2 L_0^2)$, where $F_\text{FP}$ is the finesse of the Fabry-Perot cavity and $L_0$ is the length of the interferometer arm.

\subsection{Approximation when using large-scale beam form factors.}
The denominator in Eq. (\ref{PSDGeneral}) motivates the definition of two characteristic scales for $k$. These are the Debye scale
\begin{equation}
k_D\equiv \frac{1}{ l_D},
\end{equation}
and the thermal diffusion scale,
\begin{equation}
 k_{\rm th}\equiv \frac{1}{ l_{\rm th}}=\sqrt\frac{\omega}{ D},
\end{equation}
where $l_{\rm th}$ is the characteristic distance of thermal diffusion over the time $1/\omega$.

If the characteristic lengthscale of $q(\vec{r})$ is much greater than either $l_D$ or $l_{\rm th}$, then the integrand of Eq.~(\ref{PSDGeneral}) will only be significant for small values of $k$, so that the term $Dk^2$ can be neglected in the denominator. In this approximation, the noise is then given by
\begin{equation}
 S_{\delta z}(f)\simeq \frac{4D\alpha^2 N_0}{\omega^2+(D / l_D^2)^2}\int d^3 r~\left|\nabla q\right|^2.
 \label{adiabatic}
\end{equation}
In the case of test masses with silicon substrates, the conditions for this approximation are satisfied at higher temperatures where the Debye length decreases significantly.

\section{\label{sec:level1}TCCR Noise Computation}
In this section we will compute $S_{\delta z}$ in two distinct cases:
\begin{enumerate}[label=(\Alph*)]
    \item Neglecting the standing wave by using the same beam form factor as was used in Bruns et al.~\cite{TCCRPaper}, and considering the dominant contribution to TCCR noise to come from diffusion of charge carriers in the transverse plane along the field gradient of the Gaussian beam. This will allow us to make a direct comparison of our results with those of \cite{TCCRPaper}.
    \item Including the standing wave, and separately considering the contributions to TCCR noise from diffusion of charge carriers along the field gradient of the standing wave and diffusion in the transverse plane.
\end{enumerate}
The noise power spectral density as derived by Bruns et al. will be denoted by $S^{\text{Lang}}_{\delta z}$, and our own results will be similarly labeled with superscripts.

\subsection{TCCR Noise from Transverse Diffusion, Neglecting Standing Wave} \label{transverse_results_section}
Following \cite{TCCRPaper}, we will first consider the case where TCCR noise is created exclusively by charge carriers' random walks in directions transverse to the laser beam. The resulting noise power spectral density will be denoted by $S^{\text{Tr}}_{\delta z}$. The beam form factor is given by
\begin{equation}
    q(\vec{r})=\frac{1}{\pi r_{0}^{2}} ~ \text{exp}\left(\frac{-r_\perp^{2}}{r_{0}^{2}}\right),
\end{equation}
where in Cartesian coordinates $r_\perp=\sqrt{x^2 + y^2}$ and the direction of propagation of the beam is along the $z$ axis. The form factor is normalized such that when integrating over the full volume of the input test mass we obtain
\begin{equation}
    \int_V d^3r~q(\vec{r})=L,
\end{equation}
where $L$ is the length of the input test mass along the direction of  beam propagation.
Using this form factor and substituting it into Eq.~(\ref{adiabatic}), we get
\begin{equation}
     S_{\delta z}^{\text{Tr}}(f) \simeq \frac{4D\alpha^{2} N_0 L}{\pi r_{0}^{4}}
    ~\Bigg[\frac{1}{\omega^{2} + (D / l_D^2)^2}\Bigg],
    \label{transversePSD_approx}
\end{equation}
in the limit where $r_0\gg l_D, l_{\rm th}$. Using Eq.~(\ref{PSDGeneral}) and performing some algebra, one can show that in the most general case the noise can be expressed as
\begin{equation}
    S_{\delta z}^{\text{Tr}}(f)  = \frac{4D\alpha^{2}N_0 L}{\pi r_{0}^{4}}
    ~\Bigg[\frac{1}{\omega^{2} + (D / l_D^2)^2}\Bigg]~ \xi \left(\frac{r_0^2}{2 l_D^2},\frac{r_0^2}{ 2 l_{\rm th}^2}\right),
    \label{generalformula}
\end{equation}
where the function $\xi$ is given by 
\begin{eqnarray}
    \xi(a,b)&\equiv& \int_0^{\infty}\frac{(a^2+b^2)x~\text{exp}(-x)}{ (x+a)^2+b^2}~ dx\\
     &=&\frac{|z|^2}{ 2i b}\left[z~\text{exp}(z)\Gamma(0,z)-\bar{z}~\text{exp}(\bar{z}) \Gamma(0,\bar{z})\right] \label{xi_expression}
\end{eqnarray}
where $z=a+i b$, and $\Gamma(s, z)$ is the incomplete Gamma-function. One can see that $\xi \simeq 1$ when either of its arguments is much greater than $1$, thus recovering the approximated form of Eq.~(\ref{transversePSD_approx}).

\subsection{TCCR Noise with Inclusion of Standing Wave} \label{SW_FDT_results}
Now we will compute TCCR noise with the correction of including a standing wave in the beam form factor. We will derive and compare two separate components of the noise: the first, $S_{\delta z}^\text{SW-Z}$, will only consider diffusion in the propagation direction along the standing wave, while the second, $S_{\delta z}^\text{SW-Tr}$, will exclusively consider diffusion in the transverse plane. Examining the diffusion directions separately will allow us to see that the dependence of TCCR noise on both types of diffusion is strongly temperature and frequency dependent, and that the standing wave effects dominate at higher temperatures.  The beam form factor is
\begin{equation}
    q(\vec{r})=\frac{2}{\pi r_{0}^{2}} ~ \text{exp}\left(\frac{-r_\perp^2}{r_{0}^{2}}\right) ~ \text{sin}^{2}(k_z z),
\end{equation}
where $k_z$ is the angular wave number corresponding to the wavelength of the laser light. 

In the case of $S_{\delta z}^\text{SW-Z}$, it is straightforward to get an analytical expression for the noise by solving for $\delta\rho$ using Eq.~(\ref{deltarho1}) and then substituting $\delta\rho$ into the last line of Eq.~(\ref{intrelation}) and time-averaging. We consider diffusion along the standing wave, and can thus use $\nabla \approx \hat{z} ~ \partial/\partial z$. From Eq.~(\ref{generalized_force}) we obtain a generalized force in the $z$ direction,
\begin{equation}
    F_z = \frac{2F_{0}\alpha k_z }{\pi r_{0}^{2}} ~ \text{exp}(i\omega t)   ~ \text{exp}\left(\frac{-r_\perp^2}{r_{0}^{2}}\right) ~ \text{sin}(2k_z z).
    \label{f_zi}
\end{equation}
The continuity equation can then be written as
\begin{equation}
    \left(\frac{\partial}{ \partial t}+\frac{D}{ l_D^2} -D\frac{\partial^{2}}{\partial z^{2}}\right)\delta\rho=-\rho_{0}\mu\frac{\partial F_z}{\partial z}.
    \label{continuityeqn_SW}
\end{equation}

\begin{figure*}
    \centering
    \includegraphics[width=\textwidth]{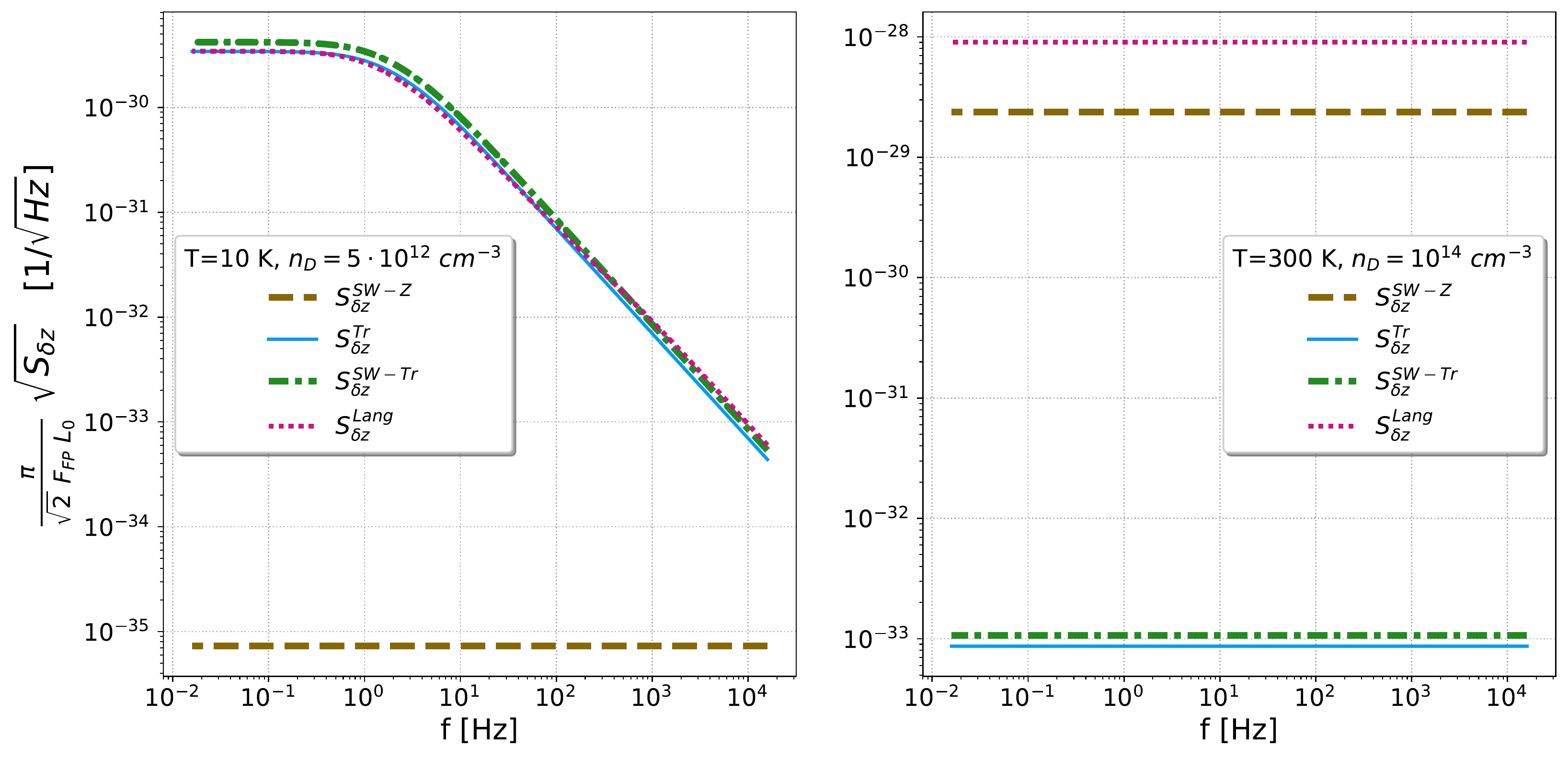}
    \caption{Amplitude spectral densities of TCCR noise are plotted above, using parameters from the ET low-frequency detector. (Left): T=10 K, $n_D = 5\times 10^{12}$ cm$^{-3}$, the lowest temperature and donor concentration proposed for ET. The noise from transverse diffusion dominates over that of propagation-direction diffusion along the standing wave. The FDT noise amplitude proportional to $\sqrt{S_{\delta z}^\text{SW-Tr}}$ is up to a factor of $\sqrt{2}$ higher than the Langevin expression. (Right): T=300 K,$~n_D$=$10^{14}$ cm$^{-3}$. Plot is reasonably representative of all temperatures from 77 K to 300 K and $n_D$ from $10^{14}$ to $10^{18}$ cm$^{-3}$, see Sec. \ref{comparison_section}. The standing wave FDT noise amplitude proportional to $\sqrt{S_{\delta z}^\text{SW-Z}}$ dominates over the transverse component and is an order of magnitude below the Langevin result, and the Langevin and FDT methods disagree by several orders of magnitude when considering identical beam form factors.}
    \label{PSDPlot}
\end{figure*} 

By inspection of the functional form of $F_z$ in Eq.~(\ref{f_zi}) and using Eq.~(\ref{continuityeqn_SW}), we consider solutions for $\delta\rho$ of the form
\begin{equation}
    \delta\rho(\vec{r},t) = C^\prime ~  \text{exp}(iwt) ~ \text{exp}\left(\frac{-r_\perp^2}{r_{0}^{2}}\right) ~ \text{cos}(2k_z z),
\end{equation}
with $C^\prime$ being a pre-factor which we can solve for to find
\begin{multline}
    \delta\rho(\vec{r},t) = -\frac{4 \rho_{0}\mu F_{0}  \alpha k_z ^{2}}{\pi r_{0}^{2}} ~ \text{exp}\left(\frac{-r_\perp^2}{r_{0}^{2}}\right)  ~ \text{cos}(2k_z z) \cdot \\ \frac{\text{exp}(iwt-i\phi)}{\left[\omega^{2}+(4Dk_z ^{2}+D/l_D^{\;2})^{2}\right]^{1/2}},
    \label{deltarho_SW}
\end{multline}
where the phase shift $\phi$ is given by
\begin{equation}
    \phi = \text{tan}^{-1}\Big(\frac{\omega}{4Dk_z ^{2}+D/l_D^{\;2}}\Big).
\end{equation}
Substituting Eq.~(\ref{deltarho_SW}) into the last line of Eq.~(\ref{intrelation}), time-averaging, and using Eq.~(\ref{S_Q_general}) along with the identity $\text{sin}(\text{tan}^{-1}(x))=|x|/(1+x^{2})^{1/2}$ gives us
\begin{equation}
    S_{\delta z}^\text{SW-Z} = \frac{4Dk_z ^2\alpha^{2}N_{0}L}{\pi r_{0}^{2}}
     ~ \Bigg[\frac{1}{\omega^{2} + \big(4Dk_z ^2 + D/l_D^{\;2}\big)^{2}}\Bigg].
    \label{standingwavePSD}
\end{equation}
Note that Eq.~(\ref{standingwavePSD}) is valid at all temperatures.

Next, it is important to note that, somewhat surprisingly, the presence of the standing wave also enhances the noise due to diffusion in the direction transverse to the beam. This can be seen as follows: for the derivation of $S_{\delta z}^\text{SW-Tr}$ we are considering diffusion only in the transverse direction, meaning $\nabla \approx \hat{x}~\partial/\partial x +  \hat{y}~\partial/\partial y$. The computation is essentially identical to that in Sec. \ref{transverse_results_section}, except that we now integrate over an additional factor of $4 ~ \text{sin}^4(k_z z)$ in Eq.~(\ref{intrelation}):
\begin{equation}
    S_{\delta z}^\text{SW-Tr}=\frac{3}{2}S_{\delta z}^\text{Tr}.
    \label{SW-TR_expression}
\end{equation}

\section{Comparison of TCCR Noise Amplitudes}\label{comparison_section}
In this section we will compare expressions for the amplitude spectral density of TCCR noise derived using the Langevin and FDT methods, and we will quantify our results with parameters from the ET low-frequency detector. All of the different expressions for TCCR noise are also plotted in Fig. \ref{PSDPlot} at temperatures of 10~K and 300~K. As described below, the 300~K plot is largely representative of a range of temperatures from 77~K to 300~K, regardless of semiconductor doping. All numerical quantities used in the figure and in this section have been taken from Tables \RomanNumeralCaps{1} and \RomanNumeralCaps{2} of \cite{TCCRPaper} unless otherwise noted.

At 10 K, the FDT-derived $(S_{\delta z}^{\text{SW-Tr}})^{1/2}$ is up to a factor of roughly $\sqrt{2}$ times greater than $(S_{\delta z}^\text{Lang})^{1/2}$, which has been numerically integrated. At higher temperatures, we can compare closed-form expressions of the noise. Remember that any comparison regarding $ S_{\delta z}^\text{Tr}$ will also apply to $ S_{\delta z}^\text{SW-Tr}$ up to a factor of 3/2. To highlight the significance of our FDT method, we can compare approximated closed-form expressions for $(S_{\delta z}^\text{Lang})^{1/2}$ and $(S_{\delta z}^\text{Tr})^{1/2}$, both of which entirely neglect the standing~wave:
\begin{equation}
     \sqrt{\frac{S^{\text{Tr}}_{\delta z}}{S^{\text{Lang}}_{\delta z}}} \simeq 2 ~ \frac{l_D}{r_{0}}.
     \label{Tr_Lang_ratio}
\end{equation}
The beam waist radius $r_0$ for ET is expected to be 0.09~m. At room temperature of 300 K and for moderately doped silicon with a donor concentration $n_D$ of $10^{14}$ $\text{cm}^{-3}$, the Debye length $l_D$ is 4.33$\times 10^{-7}$ m. Under these conditions, $(S_{\delta z}^{\text{Tr}})^{1/2}$ is roughly 5 orders of magnitude smaller than ($S_{\delta z}^\text{Lang})^{1/2}$. At 300 K and with a high $n_D$ of $10^{18}$~$\text{cm}^{-3}$, $l_D$ decreases to 2.26$\times 10^{-8}$~m. This increases the difference in the amplitudes to almost 7 orders of magnitude. The maximum and minimum values of the Debye length at other temperatures down to 77~K are within an order of magnitude of those at 300~K, meaning the disagreement shown in Eq.~(\ref{Tr_Lang_ratio}) is relatively consistent over the aforementioned temperature and doping ranges.

Next, we can extend our comparisons to $(S_{\delta z}^\text{SW-Z})^{1/2}$. At low temperatures around 10 K, the effects of diffusion along the standing wave are negligible. However, from 77~K to 300~K they are significant. We can compare closed form expressions in this temperature regime, and can ignore $\omega$ since the noise at these temperatures is constant over the whole operating frequency range of third-generation interferometers:
\begin{equation}
     \sqrt{\frac{S^{\text{SW-Tr}}_{\delta z}}{S_{\delta z}^{\text{SW-Z}}}} \simeq  \frac{4 ~  k_z ^2 ~  l_D^{\;2}+1}{k_z ~  r_{0}} \sqrt{\frac{3}{2}},
     \label{ratio_FDT_PSDs}
\end{equation}
\begin{equation}
     \sqrt{\frac{S_{\delta z}^\text{Lang}}{S_{\delta z}^\text{SW-Z}}} \simeq \frac{4 ~  k_z ^2 ~  l_D^{\;2}+1}{2 ~  k_z  ~  l_D}.
     \label{ratio_SW_Bruns_PSD}
\end{equation}
ET plans to use a 1550 nm wavelength laser, and given this fact we can see from Eq.~(\ref{ratio_FDT_PSDs}) that the noise amplitude from standing wave diffusion dominates over that of transverse diffusion by around 4 to 6 orders of magnitude, depending on doping. From Eq.~(\ref{ratio_SW_Bruns_PSD}) we can see that the noise amplitude we derive for standing wave diffusion is roughly 2 to 6 times smaller than the original prediction of Bruns et al. Again, these comparisons hold between 77~K and 300~K and for $n_D$ from $10^{14}$~to~$10^{18}$~cm$^{-3}$.

Lastly, $(S_{\delta z}^\text{SW-Z})^{1/2}$ scales in such a way that the amplitude plotted in Fig.~\ref{PSDPlot} at 300~K for $n_D=10^{14} ~\text{cm}^{-3}$ is within an order of magnitude of its amplitude at all other temperatures down to 77~K and $n_D$ up to $10^{18}$~cm$^{-3}$. Since the ET strain sensitivity goal is $3\times 10^{-25}$/$\sqrt{\text{Hz}}$, this indicates that TCCR noise should not be a limiting noise source at any temperature from 10~K to 300~K and at any feasible semiconductor doping concentration, and this finding should apply to CE and LIGO Voyager as well.

\section{Conclusion}
Semiconductor optics will provide the next generation of gravitational-wave interferometers with the potential for significant noise reduction due to their desirable thermal properties, as long as these new semiconductor materials do not introduce any significant new forms of noise. In this paper we have used the Fluctuation-Dissipation theorem to compute the power spectral density of TCCR noise, a noise source related to the thermal fluctuation of charge carriers in semiconductor materials. TCCR noise was also investigated previously by Bruns et al. \cite{TCCRPaper}, however significant differences in our approach compared with that of Bruns et al. are (1) our use of the FDT instead of an assumption of Langevin forces characterized by Poissonian charge carrier density fluctuations, and (2) our inclusion of a standing wave along the direction of propagation in the laser beam form factor.

 There are several important differences between our results and those of Bruns et al., which we quantify using parameters from Einstein Telescope. At temperatures around 10 K and for charge carrier donor concentrations of $10^{12}$ cm$^{-3}$, the FDT approach predicts that the dominant contribution to TCCR noise is diffusion along the transverse component of the beam, and that the transverse noise amplitude is around $\sqrt{2}$ times greater than what would be predicted using the Langevin approach, depending on the frequency range. The differences in our results become more significant at higher temperatures: from 77 K to 300 K and for doping concentrations from $10^{14}$ to $10^{18}$ cm$^{-3}$, the FDT method finds a noise amplitude that is 5 to 7 orders of magnitude below the Langevin prediction when using the same laser beam form factor that neglects the standing wave contribution; noise from diffusion of charge carriers along the previously neglected standing wave dominates over the noise from transverse diffusion by 4 to 6 orders of magnitude; and ultimately in this temperature regime we derive an amplitude of TCCR noise that is up to 6 times smaller than previous claims.

Despite the differences in our results and those of Bruns et al., the amplitude spectral densities we derive for TCCR noise are still several orders of magnitude below the strain sensitivity of any next-generation ground-based experiments. Thus we conclude that, at both cryogenic and room temperatures as well as for both lightly and heavily doped silicon, TCCR noise should not be a limiting noise source for next-generation gravitational-wave interferometers such as Cosmic Explorer, Einstein Telescope, or LIGO Voyager. Additionally, we strongly recommend that future noise studies make use of the standing wave in the beam form factor, as it can significantly enhance many different types of noise.

\section*{Acknowledgements}
We thank Debayan Mitra, Sergey Vyatchanin, Ashley Bransgrove, and Will Farr for their insightful comments and discussion. The Flatiron Institute is a division of the Simons Foundation, supported through the generosity of Marilyn and Jim Simons.
\nocite{*}

\bibliography{aTCCRPaper}

\end{document}